# DESIGN OF A REFORMED ARRAY LOGIC BINARY MULTIPLIER FOR HIGH-SPEED COMPUTATIONS

SAKIB MOHAMMAD

*School of Electrical, Computer, and Biomedical Engineering, Southern Illinois University Carbondale, Carbondale, IL 62901, U.S.A.*
*sakib.mohammad@siu.edu*

THEMISTOKLIS HANIOTAKIS

*School of Electrical, Computer, and Biomedical Engineering, Southern Illinois University Carbondale, Carbondale, IL 62901, U.S.A.*
*haniotak@engr.siu.edu*

Binary multipliers have long been a staple component in digital circuitry, serving crucial roles in microprocessor design, digital signal processing units and many more applications. This work presents a unique design for a multiplier that utilizes a reformed-array-logic approach to compute the product of two unsigned binary numbers. We employed a multiplexer and a barrel shifter to multiply partial products in a single clock cycle to speed up the traditional array logic. In addition, we have employed a combination of Carry Save Adders (CSA) and Ripple Carry Adders (RCA) to accumulate the partial products instead of using standalone RCAs to speed up the multiplication process further. Finally, we have demonstrated our design to perform multiplication of two 16-bit unsigned binary numbers on Cadence Virtuoso. Our design is modular and can be scaled up or down to accommodate the multiplication of any n-bit unsigned numbers.

*Keywords*: Binary multiplier, array logic, barrel shifter.

## 1. Introduction

A binary multiplier is an electronic circuit that utilizes digital logic to multiply numbers belonging to binary number system. Numerous strategies exist to realize a digital multiplier, most of which involve allocation of a group of numbers called as partial products, and subsequently summing them to obtain the final product [1, 2]. The most common algorithm follows a shift and add method and can be broken down into a series of steps. First, we multiply all the multiplicand bits by the rightmost bit of the multiplier (known as Least Significant Bit or LSB), resulting in the generation of the first partial product. Afterwards, we shift one bit to the left proceeding to the next multiplier bit repeat step 1 to generate the next partial product. This process is continued until the multiplier bits are exhausted. Finally, the partial products are accumulated to obtain the final product.

The shift-and-add approach is a common multiplication method that computers have long used to multiply small integers [3]. The process involves the contribution of every multiplier bit to generate partial products from the multiplicand. When a multiplier bit is 1, the corresponding partial product is identical to the multiplicand bits. Conversely, when a multiplier bit is 0 the partial product is also 0. The total number of partial products depends on the number of bits present in the multiplier. Once they have all been generated, they are shifted by specific counts. The first partial product remains unchanged, the second one is shifted one position to the left, the third one is shifted two positions to the left and so on. Finally, they all are summed together. For example, consider the multiplication of a 4-bit binary number with another 4-bit binary number, resulting in the generation and addition of four sets of partial products, ultimately yielding an 8-bit product. The problem with this method is that we have to generate and then add a handful of partial products, which is slow and tedious. Contemporary designs apply modified versions of this logic such as the Baugh-Wooley or Wallace Tree or use Array Logic [4-7].

Our design leverages the simplicity of shift and add logic but modifies it to speed up the multiplication process. We employed a multiplexer (or mux) and a barrel shifter (a digital circuit capable of shifting a data word by any specified bit position in a single clock cycle) to generate multiple numbers of partial products in a single clock cycle. In addition to reducing the total number of steps, we kept our overall algorithm simple and straightforward.

This paper is organized as follows: section 2 presents the materials and methods of our work. Results are elaborated in section 3. Closing remarks to this work and some future directions are presented in section 4.

## 2. Materials and Methods

Our workflow follows this trajectory: we developed our algorithm at first and later we implemented a 16*16-bit unsigned [8] multiplier on Cadence Virtuoso schematic editor using the said algorithm. We utilized Cadence Virtuoso's 45nm library to implement our design.

### 2.1. *Workflow*

Our main goal is to reduce the number of partial products generated for multiplication. To achieve this, n-bits (n>1) of the multiplier are considered to generate the partial products instead of a single bit. For example, a 6*6-bit multiplication can be performed by taking 3-bits of the multiplier from the rightmost side to generate a partial product which is equivalent to 3 partial products at each cycle instead of generating 6 partial products for each bit of the multiplier in 6 cycles. Therefore, 2 partial products as opposed to 6 are generated, reducing the total number of cycles of the traditional shift-and-add algorithm. This novel logic would require a few adders, a multiplexer (MUX), and a barrel shifter.

Fig. 1 depicts the working principle of the multiplication scheme with the multiplicand "A" and multiplier "B". Multiplication of any 3-bit combination with the multiplicand results in 8 possible outcomes. One of the possibilities would be the partial product depending on the bit-series of the multiplier. Our design realizes this concept in two steps. First, the initial versions of the partial products are generated using adders (i.e., initial adders) and are made available at MUX. Later, the MUX selects the correct partial product and delivers it to the barrel shifter. The barrel shifter shifts the initial partial products by a specified amount to generate the complete partial products. Finally, a big adder in the design referred to as the central adder, adds all the partial products sequentially. There are some shift registers available after the central adder that help sort the final product in the output registers.

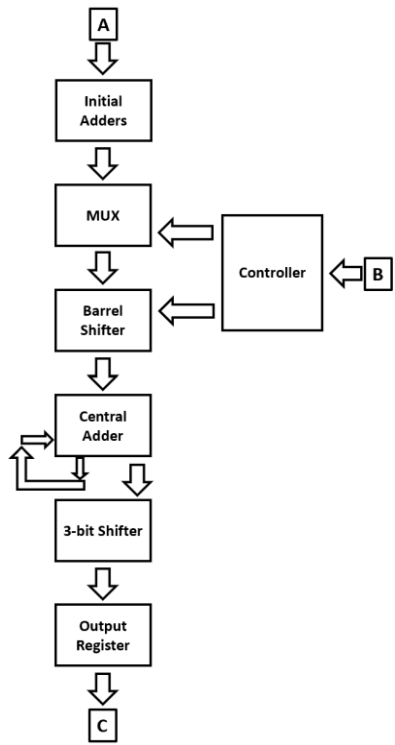

Fig. 1. Working principle of the multiplier.

Let us assume "A" and "B" are two 16-bit unsigned binary numbers, "A" and "B" being the multiplicand and multiplier, respectively. Considering 3 bits of "B" from the rightmost side, 6 partial products would be generated ("B" is padded with two zeros after the leftmost bit to make it an 18-bit number). These partial products can be anything from 0 to 7*A (since we are taking 3-bits of "B" and multiplying "A" to generate the partial products, so all possible combination of them will be between 0 (binary 000) to 7 (binary 111)) Initially, the binary equivalents of 3*A, 5*A, and 7*A are generated and will be used in the following steps to produce the final versions. The initial partial products and "A" are sent to the multiplexer. Depending on the 3 Least Significant Bits (LSB) of B, any of the A, 3*A, 5*A, or 7*A would be selected by the multiplexer and be sent to the barrel shifter. The barrel shifter will shift them either 0, 1, or 2-bit positions to the left to generate the desired final partial product (anything from 0 to 7*A). This process is repeated with the next 3-bits of "B" and continued until bits of "B" are depleted. Therefore, the MUX forwards an initial partial product at each clock cycle according to the 3 LSB of "B" ("B" provides the control bits). Afterwards, the barrel shifter shifts it to generate the final partial product, if necessary. The selection scheme of the MUX and the shifting process of the barrel shifter are presented in tables 1 and 2 respectively.

Table 1. Multiplexer's selection scheme for control bits of "B"

| 3-bits of "B" | Selection |
|---|---|
| 000 | 0 |
| 001, 010, 100 | A |
| 011, 100 | 3*A |
| 101 | 5*A |
| 111 | 7*A |

Table 2. Barrel shifter's shifting process for control bits of "B"

| 3-bits of "B" | Partial product produced | Shift position (to the left) |
|---|---|---|
| 000 | 0 | 0 |
| 001 | A | 0 |
| 010 | 2*A (will be produced by shifting A) | 1 |
| 011 | 3*A | 0 |
| 100 | 4*A (will be produced by shifting A) | 2 |
| 101 | 5*A | 0 |
| 110 | 6*A (will be produced by shifting 3A) | 1 |
| 111 | 7*A | 0 |

An adder adds the bits received from the shifter's output. After each addition, the resultant 3-LSB from the right will be sent to the output registers by the 3-bit shifter and the rest of the bits will be fed back to the adder to be added with the next batch of input bits. This process will be repeated until there is no input left to the adder. The registers following the 3-bit shifter have parallel shifting capabilities and will shift bits to the right so that the final product "C" is stored in the correct order.

*2.1.1. An example*

We will demonstrate the multiplication of two 6-bit unsigned binary numbers. Here, "A" is 001 101 and "B" is 111 111 and their product will be referred to as "C". In this paper, we are considering the rightmost bits of "A", "B" and "C" as the LSB.

The as designed multiplier produces the initial partial products 3*A, 5*A and 7*A using the initial adders at the start of the process. These values alongside A are available at the output of the MUX. Since the first 3-bits from the rightmost side of "B" are 111, we would need to generate the partial product 7*A for it. Therefore, the "B" controller selects 7*A from the mux which is shifted 0 times by the barrel shifted (which means it remains unchanged) to produce the first partial product. Let us call it PP1. The final three bits of "B" are also 111, leading to again the selection of 7*A through the MUX and the barrel shifter. We call it PP2. To obtain the final product, PP1 and PP2 have to be added by the central adder. At the beginning of this addition, 0 will be added with PP1 (since the adder is essentially empty at the beginning)) and 3-LSB from the result of this addition (011) and would be sent to the output. The rest of the bits (1101) from the results would be sent back to the central adder and would be added with PP2 in the next clock cycle. The result of this is 1101000 with 000 being sent to the output and 1101 being sent back to the adder. Since there are no partial products left, 101 is added with 0 and sent to the output and in the same way 1 would be sent to the output in the subsequent clock cycle. Finally, the output register would have 1101000011 representing the products of "A" and "B" and we would call it "C".

**2.2. *Parts of the multiplier***

In addition to the initial adders, multiplexer, barrel shifter, and the central adder, the multiplier has a series of registers that takes multiplier "B" and puts out 3-bits from the rightmost side at every clock cycle. We refer to it as the "B" controller [9]. Besides, there are some registers at the output to store the product. Both the "B" controller registers, and the output registers have shifting capabilities. The following sections will illustrate the designs of different parts of the multiplier.

*2.2.1. Initial adders, "B" controller, multiplexer, and barrel shifter*

2*A and 4*A can be produced by shifting "A" to the left by one and two-bit positions, respectively (padding zeros after the rightmost bit). Subsequently, 3*A and 5*A can be generated by adding "A" to the 2*A and 4*A, respectively. Shifting 3*A by one-bit position to the left yields 6*A addition of "A" to which generates 7*A. A, 3*A, 5*A, and 7*A are made available at the input of the MUX. 3-LSB of B will be used as the selection bits of the MUX. The barrel shifter (also controlled by similar 3-bits of "B") is placed after the MUX that further produces 2*A, 4*A, or 6*A, if necessary, by shifting A by 1 bit, 2-bits and shifting 3*A by 1-bit, respectively. The initial adders, MUX, and the barrel shifter generate the partial products needed for multiplication. Note that, for the implementation of our algorithm in this paper we considered 3-bits of "B" to generate the partial products and therefore all the variants of partial products (0 to 7*A) are available before the mux and the barrel shifter would seem to be redundant. However, for higher n-bits of "B" this would not be the situation, and all the possible partial products would not be available before the MUX. Hence, the combination of MUX and barrel shifter also helps preserve the modularity of our design to upscale or downscale the design for accommodating any bit numbers multiplication by taking any number of bits of the multiplier to generate the partial products.  Fig. 2 presents the scheme of the initial adders, multiplexer, barrel shifter and the B controller.

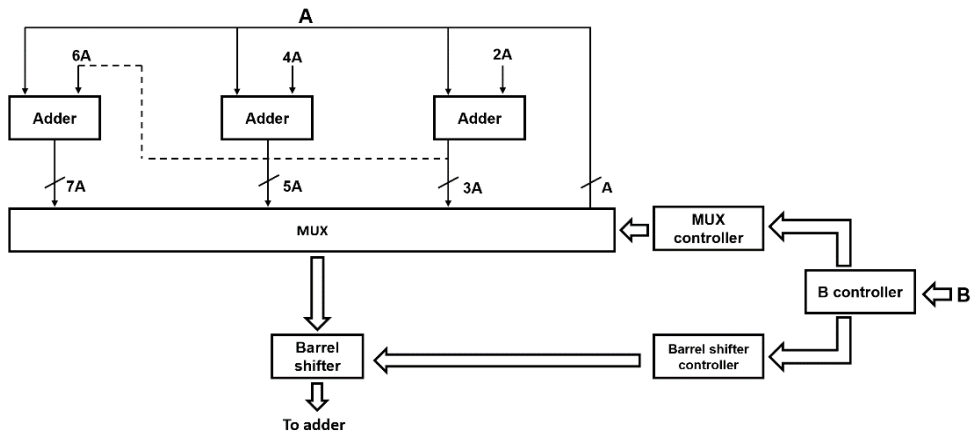

Fig. 2. The initial adders, "B" controller, multiplexer, and barrel shifter scheme of the multiplier.

The block diagrams of the B controller, MUX, and barrel shifter are depicted in figures 3(a), 3(b), and 3(c), respectively. The schematic diagrams are presented in Fig. A1, A2, and A3 in the appendix, respectively. The initial and central adder have the same design as illustrated in Fig. 4(a).

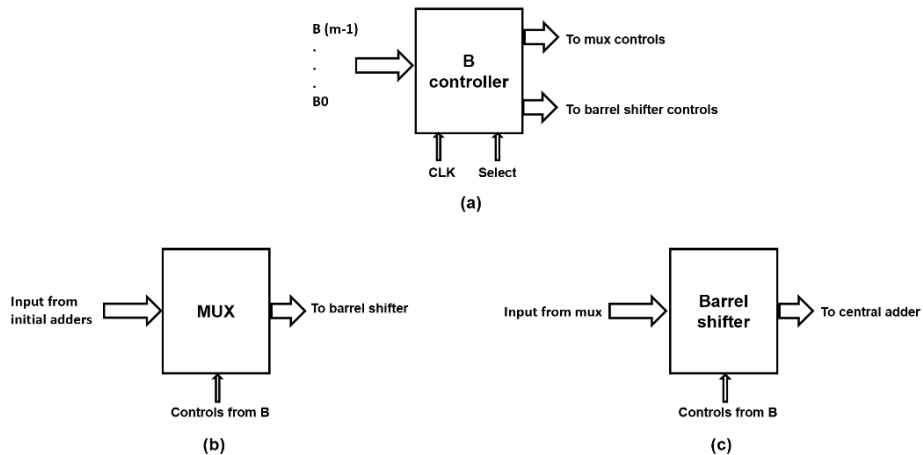

Fig. 3. Block diagram of the (a) B controller, (b) the mux and (c) the barrel shifter.

*2.2.2. Central adder and output registers*

The bits generated by the barrel shifter are made available to the central adder. The central adder is a combination of Carry Save Adders (CSAs) [10] and Ripple Carry Adders (RCAs)

[11]. The first stage is comprised of CSAs that produce a sum (S) and a carry (C). "S" and "C" are fed to RCAs in the second stage to complete the addition. From the second cycle of addition, the rest of RCA bits would be fed back to the CSA. After each complete addition, the resultant 3-LSB will be sent to 3 Parallel in Parallel Out (PIPO) shift registers, and they will send these bits to the output registers. The output registers also have shifting capabilities to store the final product of multiplication in correct order. By repeating this process until the adder is empty i.e., all the partial products have been added, the final product is available in the output registers. The block diagram of the central adder and output registers are presented in Fig. 4 (a) and (b) respectively and the schematic diagrams are presented in Fig. A4 and A5, respectively.

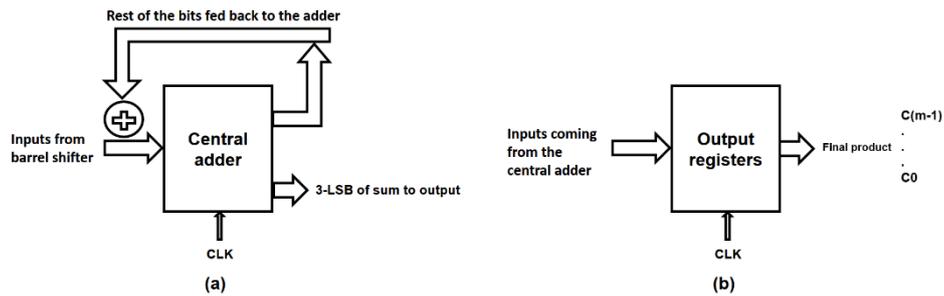

Fig. 4. (a) Central adder and (b) output registers' block diagram.

## 3. Results and Discussion

Our design has been implemented in Cadence Virtuoso schematic editor. Afterwards, the design has been run using transient analysis and it required between 500 to 800 nanoseconds for different examples. We showed an array of examples to account for both the low and high number of partial products needed to perform the multiplication. For the first example, a 16-bit unsigned binary number 0101010101010101 (A) has been multiplied by another 16-bit unsigned number 0000001011111100 (B). The numbers were chosen arbitrarily.

First, 3*A, 5*A, and 7*A were generated from "A" utilizing the initial adders and therefore A, 3*A, 5*A, and 7*A were available at the MUX. The MUX was controlled by "B". In the following step, 3-LSB of "B" were taken at each clock cycle while 7*A (for 111), 4*A (for 100), 3*A (for 011), and A (for 001) had been generated as partial products. The rest of the bits left in "B" are all 0s and for those bits, 0s were generated as partial products. All the partial products were then sent to the central adder. The adder added the partial products and sent the 3-LSB of the sum to the output registers. The rest of the bits were sent back to the adder, and they were added with the inputs of the next cycle. The central adder in this design has 25 input lines to accommodate all the partial products generated for a 16*16-bit multiplication. The final product of A*B = C = 01111110101010011010101100 is shown in Fig. 5.

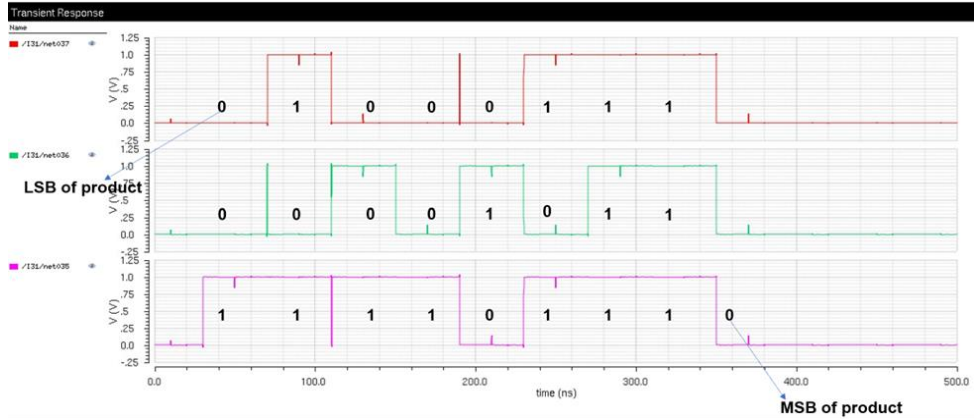

Fig. 5. Product of the first example.

In Fig. 5, the net 037 is the LSB and net 035 is the most significant bit (MSB) line when 3-bits are considered at a time from the rightmost bit. The entire process of multiplication was executed in 350 nanoseconds. It also includes the initial delay of 30 nanoseconds that was required to load "B". Therefore, the process requires 320 nanoseconds or 8 cycles since each cycle is 40 nanoseconds and is a design parameter set by the clock frequency.

For the second example, 010101010101010 (A) is multiplied by 111111111111111 (B) to produce the product 010101010101001010101010101011 (C). It took 470 - 30 = 440 nanoseconds or 11 cycles to complete. Fig. 6 illustrates the multiplication result.

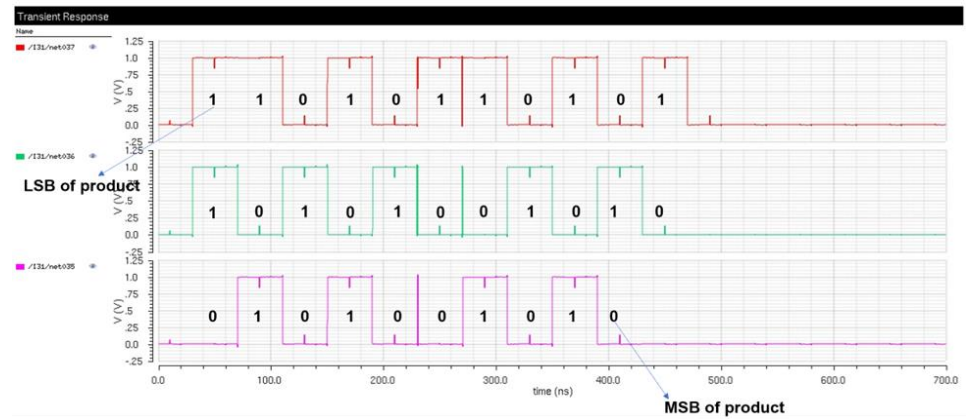

Fig. 6. Product of the second example.

Finally for the third example, we multiplied 111111111111111 (A) by 111111111111111 (B) to obtain 011111111111111100000000000000001 (C). Similar to the previous example, it took 11 cycles to finish. Fig. 7 depicts the result of multiplication.

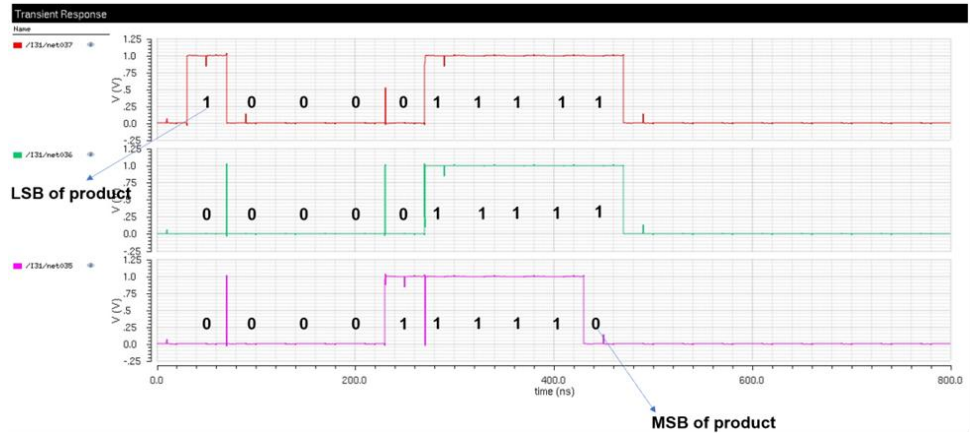

Fig. 7. Product of the third example.

It is important to observe that the total execution time is dependent on the multiplier "B" and not on the multiplicand "A" in this design. This is because "B" determines the number of partial products and that subsequently increases the addition time as well.

## 4. Conclusion

In this work, we presented a novel design for a binary multiplier that employs a reformed array logic approach for computing the product of two unsigned binary integers. By employing a barrel shifter and a multiplexer we were able to generate partial products in a lower number of steps and thus reduce the computation overhead.

The design has been evaluated by correctly demonstrating multiplication of two 16-bit unsigned numbers by taking 3-bits of the multiplier to generate the partial products. However, this design is modular and can easily be scaled up or down to accommodate various applications. The example multiplications could have been performed by using any n-bits of the multiplier without significant alternation of the design. The proposed reformed array logic binary multiplier can serve as a crucial component in microprocessor design or various DSP tasks.

We envision an extension to this design by adding zero skip mechanism. It is expected to enhance the efficiency of the multiplier further by outright skipping series of zeros in the multiplier as opposed to generating zeros as partial products. It will result in lower execution time, high throughput, and lower power consumption. A complete module will have provisions to accommodate signed numbers as well.

**Appendix A.**

The schematic diagrams of the parts of the multiplier namely the "B" controller, the MUX and its control circuitry, the barrel shifter and its controls, the central adder and the 3-bit shifter at its end and finally the output registers are presented in this section.

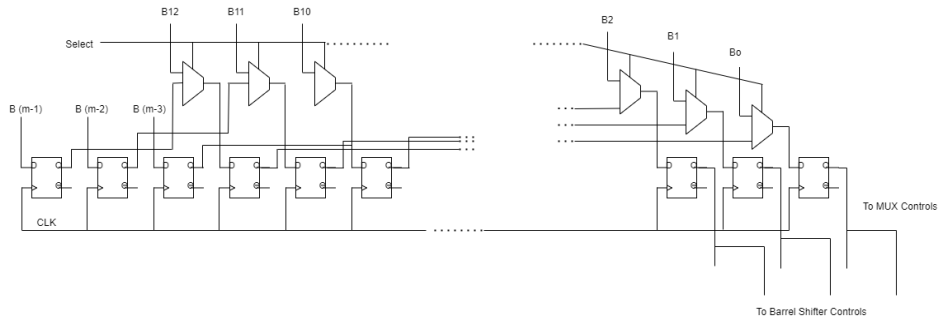

Fig. A1. The "B" controller.

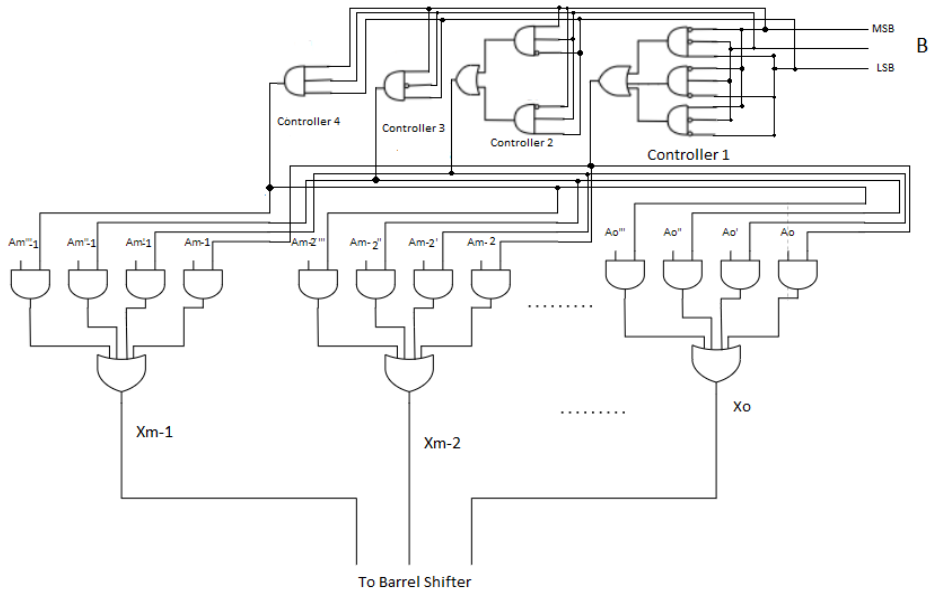

Fig. A2. The multiplexer (MUX) and its control circuit.

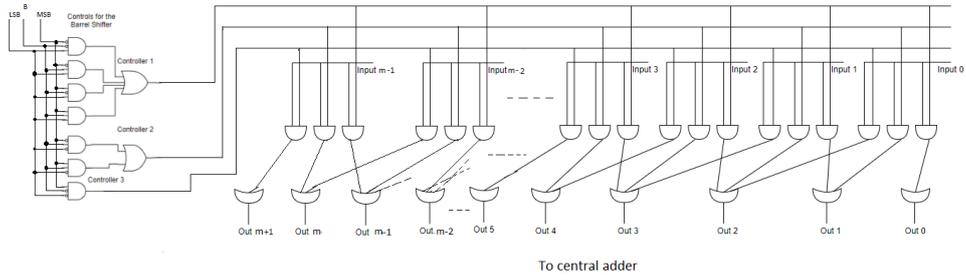

Fig. A3. The barrel shifter and its control circuitry.

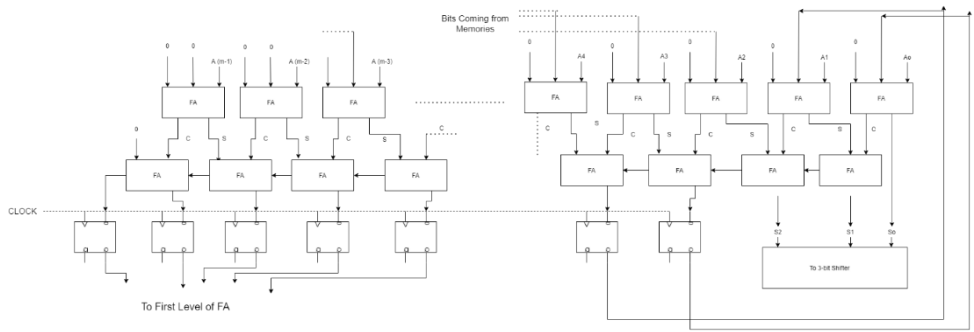

Fig. A4. The central adder and the 3-bit shifter at its end.

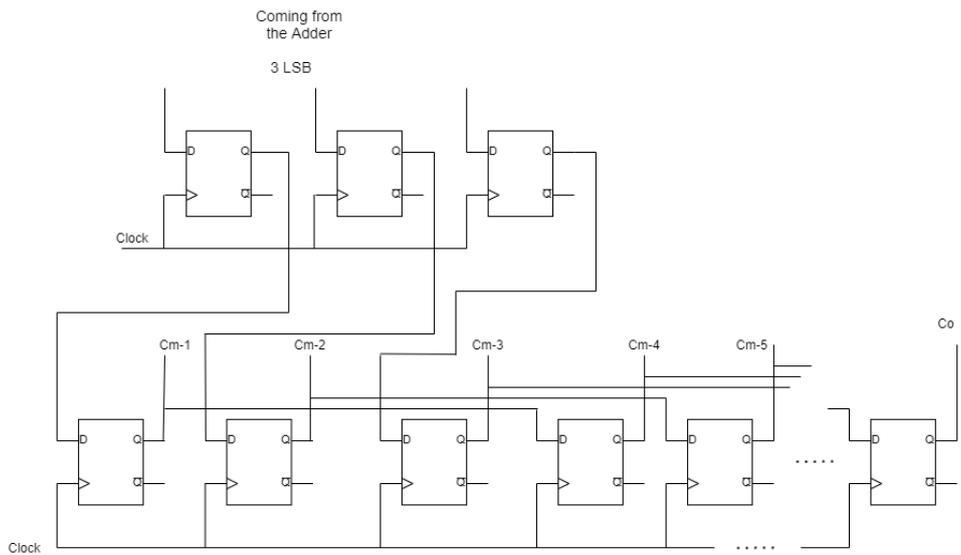

Fig. A5. The output registers that sort the final product.